\documentclass{article}

\def\p{\mbox{\boldmath$\displaystyle\mathbf{p}$}}
\def\beq{\begin{eqnarray}}
\def\eeq{\end{eqnarray}}

\begin{document}

\title{How to construct self/anti-self charge conjugate states for higher spins? }


\author{Valeriy V. Dvoeglazov\\
Universidad de Zacatecas, M\'exico\\
E-mail:vdvoeglazov@yahoo.com.mx}

\date{\empty}
\maketitle

\begin{abstract}
 We construct self/anti-self charge conjugate (Majorana-like) states for the $(1/2,0)\oplus (0,1/2)$
representation of the Lorentz group, and their analogs for higher spins within the quantum field theory. The problem of the basis rotations and that of the selection of phases in the Dirac-like and Majorana-like field operators are considered. The discrete symmetries properties (P, C, T) are studied. The corresponding dynamical equations are presented. 
In the $(1/2,0)\oplus (0,1/2)$ representation 
they obey the Dirac-like equation with eight components, which has been first introduced by Markov.
Thus, the Fock space for corresponding quantum fields is doubled (as shown by Ziino).
The particular attention has been paid to the questions of chirality and helicity 
(two concepts which are frequently confused in the literature) for 
Dirac and Majorana states. We further review several experimental consequences which follow from 
the previous works of  M.Kirchbach {\it et al.} on neutrinoless double beta decay, and G.J.Ni {\it et al.}
on meson lifetimes.
\end{abstract}



\section{Majorana-like Spinors.}



During the 20th century various authors introduced {\it self/anti-self} charge-conjugate 4-spinors
(including in the momentum representation), see, e.~g.,~\cite{Majorana,Bilenky,Ziino,Ahluwalia2}. 
Later, these spinors have been studied in Refs.~\cite{Lounesto,Dvoeglazov1a,Dvoeglazov2,Kirchbach,Rocha1}.
The authros found corresponding dynamical equations, gauge transformations 
and other specific features of them. On using $C=  -e^{i\theta} \gamma^2 {\cal K}$,
the anti-linear operator of charge conjugation (${\cal K}$ is the complex conjugation operator) we  define the {\it self/anti-self} charge-conjugate 4-spinors 
in the momentum space
$C\lambda^{S,A} ({\bf p}) = \pm \lambda^{S,A} ({\bf p})\,,\,
C\rho^{S,A} ({\bf p}) = \pm \rho^{S,A} ({\bf p})$.
The Wigner matrix is $\Theta_{[1/2]}=-i\sigma_2$\,,
and $\phi_L$, $\phi_R$ can be boosted with $\Lambda_{L,R}$ 
matrices. 

Such definitions of 4-spinors differ, of course, from the original Majorana definition in the x-representation
$\nu (x) = \frac{1}{\sqrt{2}} (\Psi_D (x) + \Psi_D^c (x))$,
$C \nu (x) = \nu (x)$ that represents the positive real $C-$ parity only. However, see~\cite{Kirchbach}, 
``for imaginary $C$ parities, the neutrino mass 
can drop out from the single $\beta $ decay trace and 
reappear in $0\nu \beta\beta $, a curious and in principle  
experimentally testable signature for a  non-trivial impact of 
Majorana framework in experiments with polarized sources."

The rest $\lambda$ and $\rho$ spinors can be defined  
in analogious way with the Dirac spinors:
\begin{eqnarray}
\lambda^S_\uparrow ({\bf 0}) = \sqrt{\frac{m}{2}}
\pmatrix{0\cr i \cr 1\cr 0},\,
\lambda^S_\downarrow ({\bf 0})= \sqrt{\frac{m}{2}}
\pmatrix{-i \cr 0\cr 0\cr 1},\,
\lambda^A_\uparrow ({\bf 0}) = \sqrt{\frac{m}{2}}
\pmatrix{0\cr -i\cr 1\cr 0},\,
\lambda^A_\downarrow ({\bf 0}) = \sqrt{\frac{m}{2}}
\pmatrix{i\cr 0\cr 0\cr 1 },\nonumber
\hspace{-20mm}\\
\end{eqnarray}
\begin{equation}
\rho^{S}_{\uparrow\downarrow} ({\bf 0}) = \mp i \lambda^A_{\downarrow\uparrow} ({\bf 0}),\,
\rho^A_{\uparrow\downarrow} ({\bf 0}) = \pm i \lambda^S_{\downarrow\uparrow} ({\bf 0})\,.
\end{equation}

Thus, in this basis with the appropriate normalization (``mass dimension")
the explicite forms of the 4-spinors of the second kind  $\lambda^{S,A}_{\uparrow\downarrow}
({\bf p})$ and $\rho^{S,A}_{\uparrow\downarrow} ({\bf p})$
are:
\begin{eqnarray}
\hspace{-10mm}\lambda^S_\uparrow ({\bf p}) &=& \frac{1}{2\sqrt{E+m}}
\pmatrix{ip_l\cr i (p^- +m)\cr p^- +m\cr -p_r },
\lambda^S_\downarrow ({\bf p})= \frac{1}{2\sqrt{E+m}}
\pmatrix{-i (p^+ +m)\cr -ip_r\cr -p_l\cr (p^+ +m)},\\
\hspace{-10mm}\lambda^A_\uparrow ({\bf p}) &=& \frac{1}{2\sqrt{E+m}}
\pmatrix{-ip_l\cr -i(p^- +m)\cr (p^- +m)\cr -p_r },
\lambda^A_\downarrow ({\bf p}) = \frac{1}{2\sqrt{E+m}}
\pmatrix{i(p^+ +m)\cr ip_r\cr -p_l\cr (p^+ +m)},
\end{eqnarray}
and
\begin{eqnarray}
\hspace{-10mm}\rho^S_\uparrow ({\bf p}) &=& \frac{1}{2\sqrt{E+m}}
\pmatrix{p^+ +m\cr p_r\cr ip_l\cr -i(p^+ +m)},
\rho^S_\downarrow ({\bf p}) = \frac{1}{2\sqrt{E+m}}
\pmatrix{p_l\cr (p^- +m)\cr i(p^- +m)\cr -ip_r},\\
\hspace{-10mm}\rho^A_\uparrow ({\bf p}) &=& \frac{1}{2\sqrt{E+m}}
\pmatrix{p^+ +m\cr p_r\cr -ip_l\cr i (p^+ +m)},
\rho^A_\downarrow ({\bf p}) = \frac{1}{2\sqrt{E+m}}
\pmatrix{p_l\cr (p^- +m)\cr -i(p^- +m)\cr ip_r}.
\end{eqnarray}

As claimed in~\cite{Ahluwalia2}, $\lambda$ and $\rho$ 4-spinors are {\it not} the eigenspinors of 
the helicity. 
Moreover, 
$\lambda$ and $\rho$ are NOT the eigenspinors of the parity, as opposed to the Dirac case 
($P=\gamma^0 R$, where $R= ({\bf x} \rightarrow -{\bf x})$).
The indices $\uparrow\downarrow$ should be referred to the chiral helicity 
quantum number introduced 
in the 60s, $\eta=-\gamma^5 h$, Ref.~\cite{SenGupta}.
The normalizations of the spinors $\lambda^{S,A}_{\uparrow\downarrow}
({\bf p})$ and $\rho^{S,A}_{\uparrow\downarrow} ({\bf p})$ have been given in the previous works.

The dynamical coordinate-space equations are:\footnote{Of course, the signs at the mass terms
depend on, how do we associate the positive- or negative- frequency solutions with $\lambda$ and $\rho$.}
\begin{eqnarray}
i \gamma^\mu \partial_\mu \lambda^S (x) - m \rho^A (x) &=& 0 \,,\,
i \gamma^\mu \partial_\mu \rho^A (x) - m \lambda^S (x) = 0 \,,
\label{12}\\
i \gamma^\mu \partial_\mu \lambda^A (x) + m \rho^S (x) &=& 0\,,\,
i \gamma^\mu \partial_\mu \rho^S (x) + m \lambda^A (x) = 0\,.
\label{14}
\end{eqnarray}
These are NOT the Dirac equation.
However, they can be written in the 8-component form. 
One can also re-write the equations into the two-component form. Thus, one obtains  
the equations of Ref.~\cite{FG}
equations. Similar formulations have been presented by M. Markov~\cite{Markov}, and by
A. Barut and G. Ziino~\cite{Ziino}. The group-theoretical basis for such doubling has been given
in the papers by Gelfand, Tsetlin and Sokolik~\cite{Gelfand}, who first presented 
the theory in the 2-dimensional representation of the inversion group in 1956 (later called as ``the Bargmann-Wightman-Wigner-type quantum field theory" in 1993).

The Lagrangian is
\begin{eqnarray}
&&{\cal L}= \frac{i}{2} \left[\bar \lambda^S \gamma^\mu \partial_\mu \lambda^S - (\partial_\mu \bar \lambda^S ) \gamma^\mu \lambda^S +
\bar \rho^A \gamma^\mu \partial_\mu \rho^A - (\partial_\mu \bar \rho^A ) \gamma^\mu \rho^A +
\bar \lambda^A \gamma^\mu \partial_\mu \lambda^A\right.\\
&&\hspace{-10mm}\left. - (\partial_\mu \bar \lambda^A ) \gamma^\mu \lambda^A +
\bar \rho^S
\gamma^\mu \partial_\mu \rho^S - (\partial_\mu \bar \rho^S ) \gamma^\mu \rho^S \right ]
 - m (\bar\lambda^S \rho^A +\bar \rho^A \lambda^S -\bar\lambda^A \rho^S -\bar\rho^S \lambda^A )\nonumber
\end{eqnarray}

The connection with the Dirac spinors has been found~\cite{Dvoeglazov1a,Kirchbach}. We can see
that the two sets are connnected by the unitary transformations, and this represents
itself the rotation of the spin-parity basis.

It was shown~\cite{Dvoeglazov1a} that the covariant derivative (and, hence, the
 interaction) can be introduced in this construct in the following way
$\partial_\mu \rightarrow \nabla_\mu = \partial_\mu - ig \L^5 A_\mu$,
where $\L^5 = \mbox{diag} (\gamma^5 \quad -\gamma^5)$, the $8\times 8$
matrix. In other words, with respect to the transformations
\begin{eqnarray}
\lambda^\prime (x)
\rightarrow (\cos \alpha -i\gamma^5 \sin\alpha) \lambda
(x)\,,\,
\overline \lambda^{\,\prime} (x) \rightarrow
\overline \lambda (x) (\cos \alpha - i\gamma^5
\sin\alpha)\,,\label{g20}\\
\rho^\prime (x) \rightarrow  (\cos \alpha +
i\gamma^5 \sin\alpha) \rho (x)\,,\,
\overline \rho^{\,\prime} (x) \rightarrow  \overline \rho (x)
(\cos \alpha + i\gamma^5 \sin\alpha)\,\label{g40}
\end{eqnarray}
the spinors retain their properties to be self/anti-self charge conjugate
spinors and the proposed Lagrangian~\cite{Dvoeglazov1a} remains to be invariant.
This tells us that while self/anti-self charge conjugate states have
zero eigenvalues of the ordinary (scalar) charge operator but they can
possess the axial charge (cf.  with the discussion of~\cite{Ziino} and
the old idea of R. E. Marshak -- they claimed the same).

Next, due to the fact that the transformations
\begin{eqnarray}
\lambda_S^\prime ({\bf p}) &=& \pmatrix{\Xi &0\cr 0&\Xi } \lambda_S ({\bf p})
\equiv \lambda_A^\ast ({\bf p}),
\lambda_S^{\prime\prime} ({\bf p}) = \pmatrix{ i\Xi &0\cr 0&-i\Xi } \lambda_S
({\bf p}) \equiv -i\lambda_S^\ast ({\bf p}),\\
\lambda_S^{\prime\prime\prime} ({\bf p}) &=& \pmatrix{ 0& i\Xi\cr
i\Xi &0\cr } \lambda_S ({\bf p}) \equiv i\gamma^0 \lambda_A^\ast
({\bf p}),\,
\lambda_S^{IV} ({\bf p}) = \pmatrix{0& \Xi\cr
-\Xi&0\cr } \lambda_S ({\bf p}) \equiv \gamma^0\lambda_S^\ast
\end{eqnarray}
with the $2\times 2$ matrix $\Xi$ defined in Ref.~\cite{Ahluwalia2},
$\Xi \Lambda_{R,L} ({\bf p} \leftarrow
{\bf 0}) \Xi^{-1} = \Lambda_{R,L}^\ast ({\bf p} \leftarrow
 {\bf 0})$\,,
and corresponding transformations for
$\lambda^A$, do {\it not} change the properties of bispinors to be in the
self/anti-self charge-conjugate spaces, the Majorana-like field operator
($b^\dagger \equiv a^\dagger$) admits additional phase (and, in general,
normalization) transformations $\nu^{ML\,\,\prime}
(x^\mu) = \left [ c_0 + i({\bf \tau}\cdot  {\bf c}) \right
]\nu^{ML\,\,\dagger} (x^\mu)$,
where $c_\alpha$ are
arbitrary parameters. The ${\bf \tau}$ matrices are defined over the
field of $2\times 2$ matrices. One can parametrize $c_0 = \cos\phi$ and ${\bf c} = {\bf n}
\sin\phi$ and, thus, define the $SU(2)$ group of phase transformations.
One can select the Lagrangian which is composed from the both field
operators (with $\lambda$ spinors and $\rho$ spinors)
and which remains to be
invariant with respect to this kind of transformations.  The conclusion
is: it is permitted the non-Abelian construct which is based on
the spinors of the Lorentz group only (cf. with the old ideas of T. W.
Kibble and R. Utiyama) .  

The Dirac-like and Majorana-like field operators can
be built from both $\lambda^{S,A} ({\bf p})$ and $\rho^{S,A} ({\bf p})$,
or their combinations. It is interesting to note that
$\left [ \nu^{^{ML}} (x^\mu) \pm C \nu^{^{ML\,\dagger}} (x^\mu) \right
]/2$ lead naturally to the Ziino-Barut scheme of massive chiral
fields, Ref.~\cite{Ziino}, if the former are composed from $\lambda^{S,A}$ spinors.
Recently, the interest to these models raised
again~\cite{Rocha1,Rocha2}.


\section{Chirality and Helicity.}



Ahluwalia~\cite{Ahluwalia2} claimed "Incompatibility 
of Self-Charge Conjugation with Helicity Eignestates and Gauge Interactions". I showed that the gauge interactions 
of $\lambda$ and $\rho$ 4-spinors are different. As for the self/anti-self charge-conjugate states
and their relations to helicity eigenstates the question is much more difficult, see below. Either we should accept that the
rotations would have physical significance, or, due to some reasons, we should not apply the equivalence
transformation to the discrete symmetry operators.
As far as I understood~\cite{Ahluwalia2} paper,  the latter standpoint is precisely his standpoint. He claimed~\cite{Ahluwalia2}: ``Just as the operator of parity in the $(j, 0) \oplus (0, j)$ representation space is independent
of which wave equation is under study, similarly the operations of charge conjugation and time
reversal do not depend on a specific wave equation. Within the context of the logical framework
of the present paper, without this being true we would not even know how to define self-/anti self
conjugate $(j, 0) \oplus (0, j)$ spinors."

Z.-Q. Shi and G. J. Ni  promote a very extreme standpoint. Namely, ```the spin states, the helicity states and the chirality states of fermions in Relativistic Quantum Mechanics ... are entirely different: a spin state is helicity degenerate; a helicity state can be expanded as linear combination of the chirality states; the polarization of fermions in flight must be described by the helicity states" (see also his Conclusion Section~\cite{Shi}). In fact, they showed experimental consequences of their statement: ``the lifetime
of RH polarized fermions is always greater than of LH ones with the same speed in flight". However, we showed that the helicity, chiral helicity and chirality
operators are connected by the unitary transformations. Do rotations have physical significance in their opinion?

M. Markov wrote long ago~\cite{Markov} {\it two} Dirac equations with  opposite signs at the mass term
$\left [ i\gamma^\mu \partial_\mu - m \right ]\Psi_1 (x) = 0$,\,
$\left [ i\gamma^\mu \partial_\mu + m \right ]\Psi_2 (x) = 0$.
In fact, he studied all properties of this relativistic quantum model (while he did not know yet the quantum
field theory in 1937). Next, he added and  subtracted these equations. What did he obtain?
\begin{equation}
i\gamma^\mu \partial_\mu \chi (x) - m \eta (x) = 0\,,\,
i\gamma^\mu \partial_\mu \eta (x) - m \chi (x) = 0\,,
\end{equation}
thus, $\chi$ and $\eta$ solutions can be presented as some superpositions of the Dirac 4-spinors $u-$ and $v-$.
These equations, of course, can be identified with the equations for $\lambda$ and $\rho$ we presented above.
As he wrote himself he was expecting ``new physics" from these equations. 
Sen Gupta~\cite{SenGupta}  and others claimed that the solutions of the equation (which follows from the general Sakurai
method of derivation of relativistic quantum equations and it may describe both massive and massless $m_1 =\pm m_2$ 
states) $\left [
i\gamma^\mu \partial_\mu - m_1 -m_2 \gamma^5 \right ]\Psi = 0$
are {\it not} the eigenstates of chiral [helicity] operator $\gamma_0 ({\bf \gamma} \cdot {\bf p})/p$
in the massless limit.
However, in the massive case the above equation has been obtained by the equivalence transformation
of $\gamma$ matrices. 
Barut and Ziino~\cite{Ziino} proposed yet another model. They considered
$\gamma^5$ operator as the operator of charge-conjugation. Thus, the charge-conjugated
Dirac equation has the different sign comparing with the ordinary formulation $[i\gamma^\mu \partial_\mu + m] 
\Psi_{BZ}^c =0$,
and the so-defined charge conjugation applies to the whole system, fermions+electromagnetic field, $e\rightarrow -e$
in the covariant derivative. The concept of the doubling of the Fock space has been
developed in Ziino works (cf.~\cite{Gelfand,Dvoeglazov5}). In their case, see above, their charge conjugate states
are at the same time the eigenstates of the chirality.

Let us analize the above statements.
The helicity operator is $\hat h=\frac{1}{2}\pmatrix{({\bf \sigma}\cdot \hat{\bf p})&0\cr
0&({\bf \sigma}\cdot \hat{\bf p})\cr }$.
However, we can do the equivalence transformation of the  helicity $\hat h$-operator by
the unitary matrix. It is known~\cite{Berg}
that one can ${\cal U}_1 ({\bf \sigma}\cdot {\bf a}) {\cal U}_1^{-1} = \sigma_3 \vert {\bf a} \vert$.
In the case of the momentum vector (${\bf n} \equiv \hat{\bf p} = {\bf p}/\vert {\bf p}\vert$) , one has
\begin{equation}
{\cal U}_1 =\pmatrix{1& p_l/(p+p_3)\cr
-p_r/(p+p_3)&1\cr},\,\,U_1 =\pmatrix{{\cal U}_1 &0\cr
0& {\cal U}_1\cr}\,,\,\,
U_1 \hat h  U_1^{-1} = 
\vert \frac{{\bf n}}{2} \vert \pmatrix{\sigma_3 &0\cr
0&\sigma_3 }.
\end{equation} 
Then, applying other unitary matrix $U_3$:
\begin{eqnarray}
&&
\hspace{-10mm}
\pmatrix{1&0&0&0\cr
0&0&1&0\cr
0&1&0&0\cr
0&0&0&1\cr } \pmatrix{\sigma_3 &0\cr
0&\sigma_3 } \pmatrix{1&0&0&0\cr
0&0&1&0\cr
0&1&0&0\cr
0&0&0&1\cr } = \pmatrix{1&0&0&0\cr
0&1&0&0\cr
0&0&-1&0\cr
0&0&0&-1\cr }
\,.
\end{eqnarray}
we transform to the basis, where helicity is equal (within the factor $\frac{1}{2}$) to $\gamma^5$, the chirality operator.

The author of~\cite{SenGupta} and others introduced the {\it chiral} helicity $\eta =-\gamma_5 h$, which is equal 
(within the sign and the factor $\frac{1}{2}$) to the well-known matrix ${\bf \alpha}$ multiplied by ${\bf n}$. Again,
$U_1 ({\bf \alpha}\cdot {\bf n}) U_1^{-1} = 
\alpha_3 \vert {\bf n}\vert$\,,
with the same matrix
$U_1$.
And applying the second unitary transformation:
\begin{equation}
U_2 \alpha_3 U_2^\dagger =
\pmatrix{1&0&0&0\cr
0&0&0&1\cr
0&0&1&0\cr
0&1&0&0\cr } \alpha_3 \pmatrix{
1&0&0&0\cr
0&0&0&1\cr
0&0&1&0\cr
0&1&0&0\cr } =
\pmatrix{1&0&0&0\cr
0&1&0&0\cr
0&0&-1&0\cr
0&0&0&-1\cr }
\end{equation}
we again come to the $\gamma_5$ matrix. The determinats are: $Det U_1=1\neq 0$, $Det U_{2,3}=-1\neq 0$. Thus, helicity, chirality and chiral  helicity are connected by the unitary transformations.

It is {\it not} surprising to have such a situation because the different helicity 2-spinors can be also 
connected {\it not only}
by the anti-linear transformation~\cite{Ryder,Ahluwalia2},
\linebreak  $\xi_h = (-1)^{1/2+h} e^{i\alpha_h} \Theta_{[1/2]} {\cal K} \xi_{-h}$, but the unitary transformation too. For instance, when we parametrize the 2-spinors as in~\cite{Dv-ff}
we obtain
\begin{equation}
\xi_\downarrow = U\xi_\uparrow = e^{i (\beta -\alpha)} \pmatrix{0& e^{-i\phi}\cr
-e^{i\phi}&0\cr }\xi_\uparrow,\,\,
\xi_\uparrow = U^\dagger \xi_\downarrow = e^{i(\alpha -\beta)} \pmatrix{0& -e^{-i\phi}\cr
e^{i\phi}&0\cr }\xi_\downarrow\,.
\end{equation}

To  say that the 4-spinor is the eigenspinor of the {\it chiral helicity}, and, at the same time, it is {\it not!}
the eigenspinor of the helicity operator (and that the physical results would depend on this) signifies the same as to say that rotations have physical significance
on the fundamental level.


\section{Charge Conjugation and Parity for $S=1$.}




Several formalisms have been used for higher spin fields, e.~g.,~\cite{BW, Weinberg}.
The $2(2S+1)$ formalism gives the equations which are in some sense on an equal footing with the
Dirac equation. For instance, for the spin-1 field the equation is
$[\gamma_{\mu\nu} p_\mu p_\nu - m^2] \Psi (x) =0$\,,
with the $\gamma_{\mu\nu}$ being the 6x6 covariantly-defined matrices.
However, it was argued later that the signs before the mass terms should be opposite 
for charged particles of positive- and negative- frequencies~\cite{Sankaranarayanan,Ahluwalia1}:
$[\gamma_{\mu\nu} p_\mu p_\nu - (\frac{i\partial/\partial t}{E}) m^2] \Psi (x) =0$\,.
Hence, Ahluwalia {\it et al.} write: "The charge conjugation operation $C$
must be carried through with a little greater care for bosons than
for fermions within [this] framework because of
$\wp_{u,v}=\pm 1$ factor in the mass term. For the $(1,0)\oplus(0,1)$ case,
at the classical level we want
\begin{equation}
C:\,
\left(\gamma_{\mu\nu}\,D^\mu_{+}\,D^\nu_{+}\,+\,m^2\right)\,u(x)=0
\,\rightarrow\,
\left(\gamma_{\mu\nu}\,D^\mu_{-}\,D^\nu_{-}\,-\,m^2\right)\,v(x)=0,\label{c}
\end{equation}
where the local $U(1)$ gauge covariant derivatives are defined as:
$
D^\mu_{+}\,=\,\partial^\mu\,+\,i\,q\, A^\mu(x)$,
$D^\mu_{-}\,=\,\partial^\mu\,-\,i\,q\, A^\mu(x)$", Ref.~\cite{Ahluwalia1}.

"These results read [Ref.~\cite{Ahluwalia2}]:
\begin{equation}
S^c_{[1]}
=
e^{i\vartheta^{c}_{[1]}}
\left(
\begin{array}{cc}
0 & \Theta_{[1]}\\
-\,\Theta_{[1]} &0
\end{array}\right)\,{\cal K}\,\equiv\, C_{[1]}\,{\cal K}\,,\,
S^s_{[1]} =
e^{i\vartheta^{s}_{[1]}}\,
\left(
\begin{array}{cc}
0&1_3\\
1_3&0
\end{array}
\right)\,=\,e^{i\vartheta^{s}_{[1]}}\,\gamma_{00}
\,. \label{cpo}
\end{equation}
Note that neither $S^c_{[1/2]}$ nor  $S^c_{[1]}$ are unitary (or even
linear)." $\Theta_{[1]}$ is the 3x3 representation of the $\Theta_{[1/2]}=-i\sigma_2$.

"For spin-$1$ ... the requirement of self/anti-self charge
conjugacy {\it cannot} be satisfied. That is, there does not exist a
$\zeta$ [the phase factors between right- and left- 3-"spinors"] that can satisfy the spin-$1$ ... requirement"
$S^c_{[1]}
\,\lambda(p^\mu)\,=\,\pm\,\lambda(p^\mu)\,\,,\quad S^c_{[1]}
\,\rho(p^\mu)\,=\,\pm\,\rho(p^\mu)$"\quad (?).
This is due to the fact that $C^2 = -1$ within this definition of the charge conjugation
operator. "We find, however, that the requirement of self/anti-self
conjugacy under charge conjugation can be replaced by the requirement
of self/anti-self
conjugacy under the operation of $\Gamma^5\,S^c_{[1]}$ [precisely, which was used by Weinberg in Ref.~\cite{Weinberg}
due to the different choice of the equation for the negative-frequency 6-"bispinors"], 
where
$\Gamma^5$ is the {\it chirality} operator for the
$(1,\,0)\oplus(0,\,1)$ representation space...

The requirement
$\left[\Gamma^5\,S^c_{[1]}\right]\,
\lambda(p^\mu)\,=\,\pm\,\lambda(p^\mu)\,\,,\quad
\left[\Gamma^5\,S^c_{[1]}\right]\,
\rho(p^\mu)\,=\,\pm\,\rho(p^\mu)$
determines $\zeta^S_\lambda\,=\,+\,1\,=\,\zeta^S_\rho$ for the self
$\left[\Gamma^5\,S^c_{[1]}\right]$-conjugate  spinors $\lambda^S(p^\mu)$ and
$\rho^S(p^\mu)$; and
$\zeta^A_\lambda\,=\,-\,1\,=\,\zeta^A_\rho$ for the anti-self
$\left[\Gamma^5\,S^c_{[1]}\right]$-conjugate
spinors $\lambda^A(p^\mu)$ and $\rho^A(p^\mu)$".

The covariant equations for $\lambda-$ and $\rho-$ objects in the $(1,0)\oplus (0,1)$
representation have been obtained~in~Ref.~\cite{Dvoeglazov1a}.
under the certain choice of the phase factors in the definition of left- and right- 3-objects.

On the quantum-field level we have to introduce the unitary operators for the charge conjugation 
and the parity in the Fock space
$U^c_{[S]} \Psi_{[S]} (x^\mu) (U^c_{[S]})^{-1} =  C_{[S]}
\Psi^\dagger_{[S]} (x^\mu)$,\linebreak
$U^s_{[S]} \Psi_{[S]} (x^\mu) (U^s_{[S]})^{-1} = \gamma^0
\Psi_{[S]} (x^{\prime^{\,\mu}})$.
For the spin $S=1/2$ they can be find in the well-known textbooks~\cite{ItzyksonZuber}.

R. da Rocha {\it et al.} write~\cite{Rocha2}:
"Now let one denotes the eigenspinors of the Dirac operator for particles and antiparticles respectively by $u_{\pm}({\bf p})$ and $v_\pm({\bf p})$.  The subindex $\pm$ regards the eigenvalues of the helicity operator $(\sigma\cdot\widehat{\bf{p}})$. The parity operator acts as $
P u_\pm({\bf p}) = +\, u_\pm({\bf p})$,\,
$P v_\pm({\bf p}) = -\, v_\pm({\bf p})$,
which implies that $ P^2= \,1$ in this case.
The action of $C$ on these spinors is given  [in textbooks~\cite{ItzyksonZuber}],
which implies that $\{C,P\}=0$, [anticommutator].

On the another hand
the parity operator $P$ acts on ELKO by
$P\lambda^{S}_{\mp,\pm} (\p)= \pm\, {i}\,
\lambda^{A}_{\pm,\mp}(\p)$,\,
$P\lambda^{A}_{\mp,\pm} (\p)= \mp \,{i}
\,\lambda^{S}_{\pm,\mp}(\p)$,
and it follows that $[C,P]=0$ [when acting on the Majorana-like states]." In the previous works of 
the 50s-60s, Ref.~\cite{NigamFoldy}
it is this case which has been attributed to the $Q=0$ eigenvalues (the truly neutral particles).
You may compare these results with those of Refs.~\cite{Ahluwalia2,Dvoeglazov2,Dvoeglazov4},
where the same statements have been done on the quantum-field level even 
at the earlier time comparing with~\cite{Rocha2}. The notation for the 4-spinors used in the cited papers is a bit different.
The acronym "ELKO" is ({\bf almost}) the synonym for the self/anti-self charge conjugated states (the Majorana-like spinors).
So, why the difference appeared in the da Rocha formulas comparing with my previous results on the classical level?
In my  papers, see, e.g., Ref.~\cite{Dvoeglazov1a,Dvoeglazov2,Dvoeglazov4}, I presented the explicite forms of 
the $\lambda-$ and $\rho-$ 2-spinors in the basis $\hat S_3 \xi ({\bf 0}) =\pm \frac{1}{2}\xi ({\bf 0})$. 
The corresponding properties with respect to the parity (on the classical level) are different:
\begin{equation}
\gamma^0 \lambda^S_{\uparrow\downarrow} (p^{\mu^\prime}) = \pm i \lambda^S_{\downarrow\uparrow} (p^\mu),\,\,
\gamma^0 \lambda^A_{\uparrow\downarrow} (p^{\mu^\prime}) = \mp i \lambda^A_{\downarrow\uparrow} (p^\mu).
\end{equation}

It is easy to find the correspondence between "the new notation", Refs.~\cite{Ahluwalia3,Rocha2}
and the previous one. Namely, $\lambda^{S,A}_{\uparrow} \rightarrow \lambda^{S,A}_{-,+}$,
$\lambda^{S,A}_{\downarrow} \rightarrow \lambda^{S,A}_{+,-}$. However, the difference
is also in the choice of the basis for the 2-spinors (!). As in Ref.~\cite{Dvoeglazov3},
Ahluwalia, Grumiller and da Rocha have chosen the well-known helicity basis (cf.~\cite{Varshalovich,Dv-ff}).
In my work of 2002 (published in 2004) I have shown that the helicity-basis 4-spinors
satisfies the same Dirac equation, the parity matrix can be defined in the similar fashion 
as in the spinorial basis (according to the Itzykson-Zuber textbook \cite{ItzyksonZuber}), but 
the helicity-basis 4-spinors are {\it not} the eigenspinors of the parity
(in full accordance with the claims made in the 4th volume of the Landau course of theoretical physics
and with the fact that $ [\hat h,\hat P]_+=0$, Ref.~\cite{BLP}). 
In this basis, the parity transformation ($\theta \rightarrow \pi -\theta$, $\phi \rightarrow \pi +\phi$)
lead to the properties:
\begin{eqnarray}
R \phi_L^- ({\bf 0}) &=& -i e^{i(\theta_2 -\theta_1)}\phi_L^+ ({\bf 0})\,,\,
R \phi_L^+ ({\bf 0}) = -i e^{i(\theta_1 -\theta_2)}\phi_L^- ({\bf 0}),\\
R \Theta (\phi_L^- ({\bf 0}))^\ast &=& -i e^{-2i\theta_2}\phi_L^- ({\bf 0})\,,\,
R \Theta (\phi_L^+ ({\bf 0}))^\ast =  +i e^{-2i\theta_1}\phi_L^+ ({\bf 0}).
\end{eqnarray}
This  opposes  to the spinorial basis, where, for instance:
$R \phi_L^{\pm} ({\bf 0}) =\phi_L^{\pm} ({\bf 0})$. Further calculations are straightforward, and, indeed, 
they can lead to $[C,P]_- =0$ when acting on the "ELKO" states, due to $[C,\gamma^5]_+=0$.

In the $(1,0)\oplus (0,1)$ representation the situation is similar. If we would like to extend
the Nigam-Foldy conclusion, Ref.~\cite{NigamFoldy} (about $[C,P]_- =0$ corresponds to the neutral particles even in 
the higher spin case (?)) then we should use the helicity basis on the classical level.
However, on the level of the quantum-field theory (the ``secondary" quantization)
the situation is self-consistent. As shown in 1997, Ref.~\cite{Dvoeglazov2,Dvoeglazov4},
we can obtain easily {\bf both} cases (commutation and anti-commutation)
on using $\lambda^{S,A}$ 4-spinors, which have been used earlier (in the basis $column(1\,\, 0)$
$column (0\,\, 1)$).

\section{Conclusions.}



We presented a review of the formalism for the momentum-space Majorana-like
particles in the $(S,0)\oplus (0,S)$ representation of the Lorentz Group. The $\lambda$ and $\rho$
4-spinors  satisfy the 8-
component analogue of the Dirac equation. Apart, they have different gauge transformations
comparing with the usual Dirac 4-spinors. Their helicity, chirality and chiral helicity properties
have been investigated in detail. These operators are connected by the given unitary
transformations. At the same time, we showed that the Majorana-like 4-spinors can be obtained
by the rotation of the spin-parity basis. Meanwhile, several authors have claimed that the
physical results would be different on using calculations with these Majorana-like spinors.
Thus, the $(S,0)\oplus (0,S)$ representation space (even in the case of $S = 1/2$) has additional
mathematical structures leading to deep physical consequences, which have not yet been explored
before.
However, several claims made by other researchers concerning with chirality, helicity, chiral
helicity should not be considered to be true until the time when experiments confirm them.
Usually, it is considered that the rotations (unitary transformations) have {\it no} any physical
consequences on the level of the Lorentz-covariant theories.

Next, we discussed the $[C,P]_{\pm}=0$ dilemma for neutral and charged particles
on using the analysis of the basis rotations and phases.

\end{document}